%% file: main.tex
\documentclass[11pt,a4paper]{article}
\pdfoutput=1
\usepackage[T1]{fontenc}
\usepackage[utf8]{inputenc} 
\usepackage{lmodern}

\usepackage{authblk}
\usepackage{graphicx}
\usepackage{amsmath, amssymb}
\usepackage{subcaption}
\usepackage{verbatim}
\usepackage{indentfirst}

\title{High-Voltage Performance Testing in LAr of the PMMA Cathode Connection for the DarkSide-20k Experiment}

\author[1]{L. Luzzi\thanks{on behalf of the DS-20k Collaboration}}
\affil[1]{University of California, 1 Shields Ave, Davis, CA 95616, United States}
\affil[ ]{\texttt{lluzzi@ucdavis.edu}}

\date{} 

\begin{document}
\maketitle

\begin{abstract}
DarkSide--20k (DS--20k) is a next--generation dual--phase liquid argon (LAr) time projection chamber (TPC) devoted to the direct--detection of dark matter. The detector is currently under construction in Hall--C at the Laboratori Nazionali del Gran Sasso, Italy, at a depth of approximately 3500 m water equivalent. The detector will instrument 49.7 t of low--radioactivity underground LAr contained within an acrylic TPC and is designed to reach a WIMP--nucleon spin--independent cross--section sensitivity down to $10^{-48}\,\mathrm{cm^{2}}$ for a WIMP mass of $0.1\,\mathrm{TeV}/c^{2}$ in a 200 tonne--year run. In DS--20k a uniform electric drift field is established in the active volume to transport ionization electrons toward the electroluminescence region, with the required high voltage delivered to the TPC cathode through a custom cable and stress--cone assembly. At University of California, Davis, a dedicated test setup was developed to reproduce the DS--20k cathode high--voltage connection in LAr, matching the local electric--field conditions. This work summarizes the results of a comprehensive test campaign validating the operation of the DS--20k cathode HV system in LAr up to $-100$ kV.
\end{abstract}



\input{TexFiles/Introduction}

\input{TexFiles/HV_DS-20k}

\input{TexFiles/Setup}

\input{TexFiles/Cryogenics}

\input{TexFiles/High_Voltage}

\input{TexFiles/Conclusions}

\clearpage

\section*{Acknowledgments}
 This report is based upon work supported by the U. S. National Science Foundation (NSF) (Grants No. PHY-2310048 and PHY-2131857)

\end{document}

%% file: TexFiles/Introduction.tex
\section{Introduction}
\label{sec:Intro}

The Global Argon Dark Matter Collaboration (GADMC) is currently assembling DarkSide-20k (DS-20k)~\cite{ds_20k} at LNGS (Italy), under an effective overburden of roughly 3500~m.w.e., with a planned operational lifetime of at least $10$~years.
DS-20k is a massive dual--phase liquid argon (LAr) time projection chamber (TPC) designed for the direct detection of Weakly Interacting Massive Particles (WIMPs). With a fiducial LAr mass of 20~t, the experiment is designed to operate in a regime close to instrumental background--free conditions, with fewer than $0.1$ background events expected in the WIMP search region over the planned exposure of $200$~tonne--years \cite{canci}.

The active volume of the TPC operates with a uniform electric field of $200$~V/cm, generated by biasing the cathode at approximately $-75$~kV via a dedicated high--voltage cable and stress--cone assembly.
The key challenges associated with this system include: delivering high voltage while minimizing the risk of electrical discharges, ensuring the stability of the stress--cone under cryogenic conditions, and achieving a reliable connection within the constrained geometry of the cathode plug.

At University of California, Davis (UC Davis), the DS--20k HV stress--cone design was tested within a closed LAr system using a dedicated setup that reproduces the same local electric--field conditions expected for the actual detector. The setup employs the same HV cable and stress--cone assembly used in the experiment and is immersed in approximately 20 liters of LAr, allowing systematic studies of the system behavior and operational procedures under HV and cryogenic conditions.
This test provides key technical and operational insights into the design, operation, and long--term reliability of the DS--20k cathode HV supply and plug system.

%% file: TexFiles/HV_DS-20k.tex
\section{Electric Fields and High--Voltage in DS--20k}
\label{sec:ds_20k}

The DS--20k TPC barrel is an octagonal structure approximately $3.6$~m high, made of high--purity acrylic (PMMA), with the anode and cathode corresponding to the top and bottom PMMA planes, respectively.
The electric field inside the TPC is established by coating the inner surfaces with Clevios$^{\text{TM}}$ conductive polymer, while field uniformity is ensured by a resistor chain supplying the appropriate potentials to the field--cage rings. A thin stainless--steel (SS) wire grid, located $10$~mm below the anode cap, separates the active volume into the drift and extraction/electroluminescence regions. The detector operates with a nominal drift field of $200$~V/cm, obtained by biasing the cathode at approximately $-75$~kV~\cite{ds_20k}.

\begin{figure}[h]
    \centering
    \begin{subfigure}[b]{0.42\textwidth}
        \includegraphics[width=1.1\textwidth]{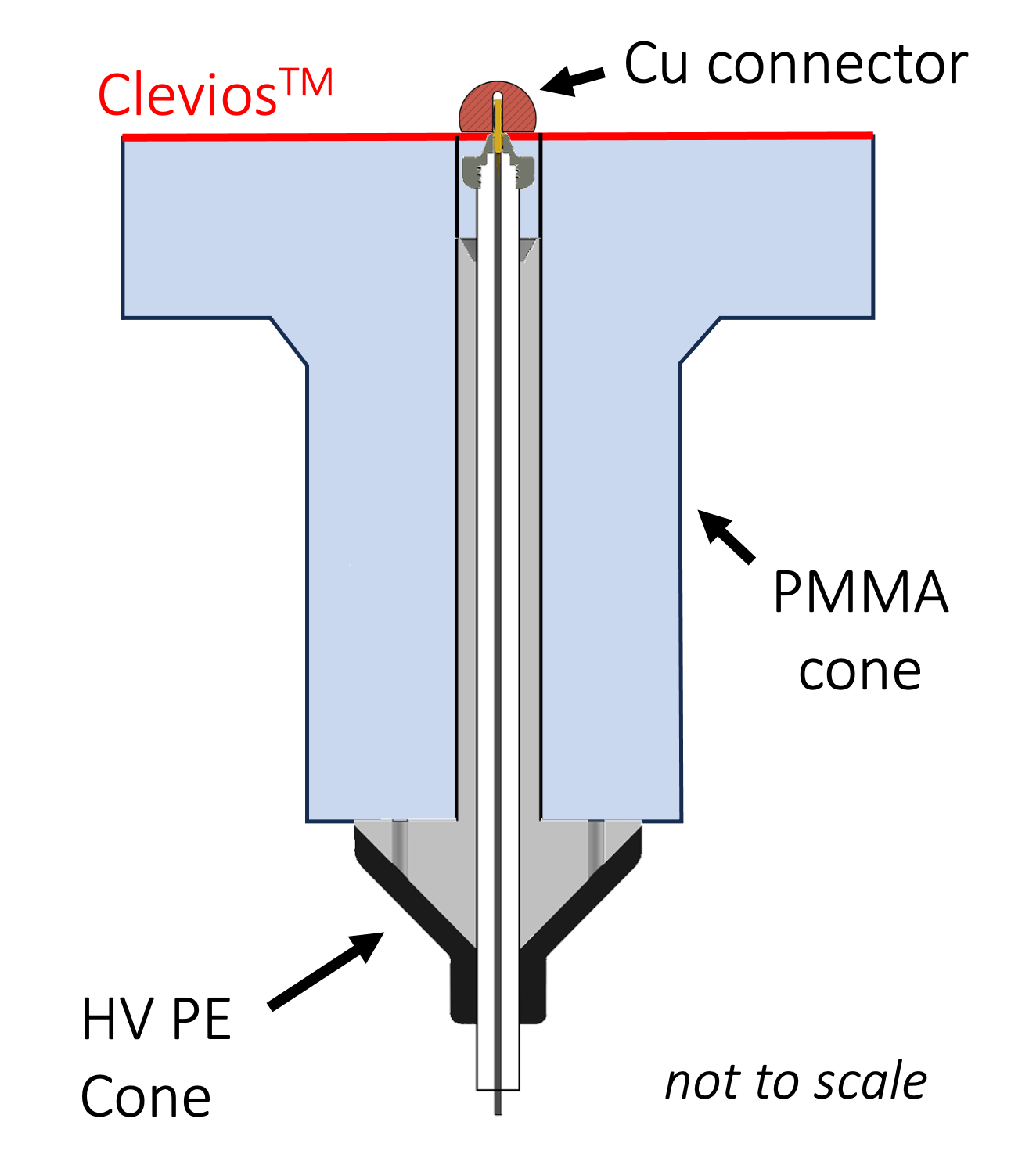}
        \caption{}
        \label{fig:cath_plug}
    \end{subfigure}
    \hspace{0.2\textwidth}
    \begin{subfigure}[b]{0.34\textwidth}
        \includegraphics[width=1.1\textwidth]{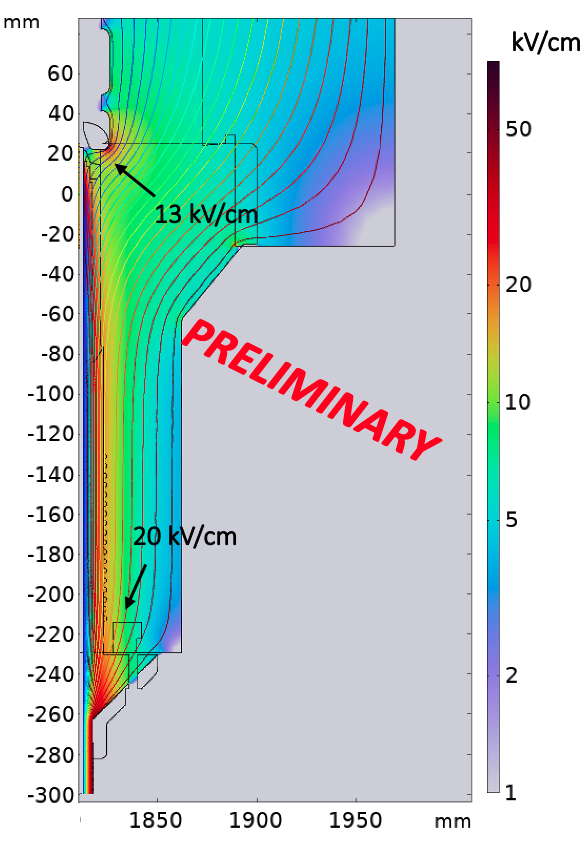}
        \caption{}
        \label{fig:comsol}
    \end{subfigure}
    \caption{Plug of the HV cable to the DS--20k cathode through the dedicated stress--cone assembly. Left: simplified schematic cross--section in the $XZ$ plane. Right: 2D \texttt{COMSOL} simulation with the cathode biased at $-75$~kV.
}
    \label{fig:cathode_plug}
\end{figure}

In DS--20k, high voltage is delivered to the cathode using a custom HV cable rated up to $150$~kV and manufactured by Dielectric Sciences (Lowell, MA, USA). The cable features a three--layer polyethylene (PE) structure, with a carbon--loaded central conductor (2 mm diameter) surrounded by a $4.6$~mm thick PE insulating layer and a $0.25$~mm thick outer carbon--loaded PE ground. The three PE layers are co-extruded during the cable manufacturing process.
The HV cable enters the DS--20k vessel and runs along the full height of the TPC, connecting to the underside of the cathode via a dedicated stress--cone assembly (Fig.~\ref{fig:cath_plug}), consisting a PE--cone plug, and a PMMA--cone bonded to the cathode PMMA window.

\noindent The PE--cone consists of an insulating PE sleeve and a carbon--loaded conductive PE outer shell, thermally fused onto the cable and fabricated by the UC Davis group. Along the cable section extending inside the stress cone toward the cathode connection, the outer ground layer is mechanically stripped by longitudinal cutting and peeling to expose the insulating PE layer.
The cable terminates at a copper connector mounted on the inner surface of the cathode, in direct electrical contact with the Clevios$^{\text{TM}}$ conductive coating.

The highest electric--field regions are located at the PE--cone/PMMA--cone interface and beneath the copper connector. A \texttt{COMSOL} study (Fig.~\ref{fig:comsol}) with the cathode biased at $-75$~kV shows that the electric field in these regions remains below the breakdown threshold in LAr ($\sim 40$~kV/cm \cite{blatter}), corresponding to a safety margin of approximately $45\%$ with respect to the nominal cathode voltage. The outer surface of the PMMA cylinder is coated with Clevios and held at ground potential. The simulation does not include spatial or time--dependent surface charge accumulation effects. Under these conditions, the cathode HV system must operate while minimizing the risk of electrical discharges, ensuring a reliable connection within a constrained geometry, and maintaining stability at liquid--argon temperature ($\sim 87$~K).

%% file: TexFiles/Setup.tex
\section{Experimental Test Setup at UC Davis}
\label{sec:setup}

\begin{figure}[h]
    \centering
    \begin{subfigure}[b]{0.31\textwidth}
        \includegraphics[width=1.1\textwidth]{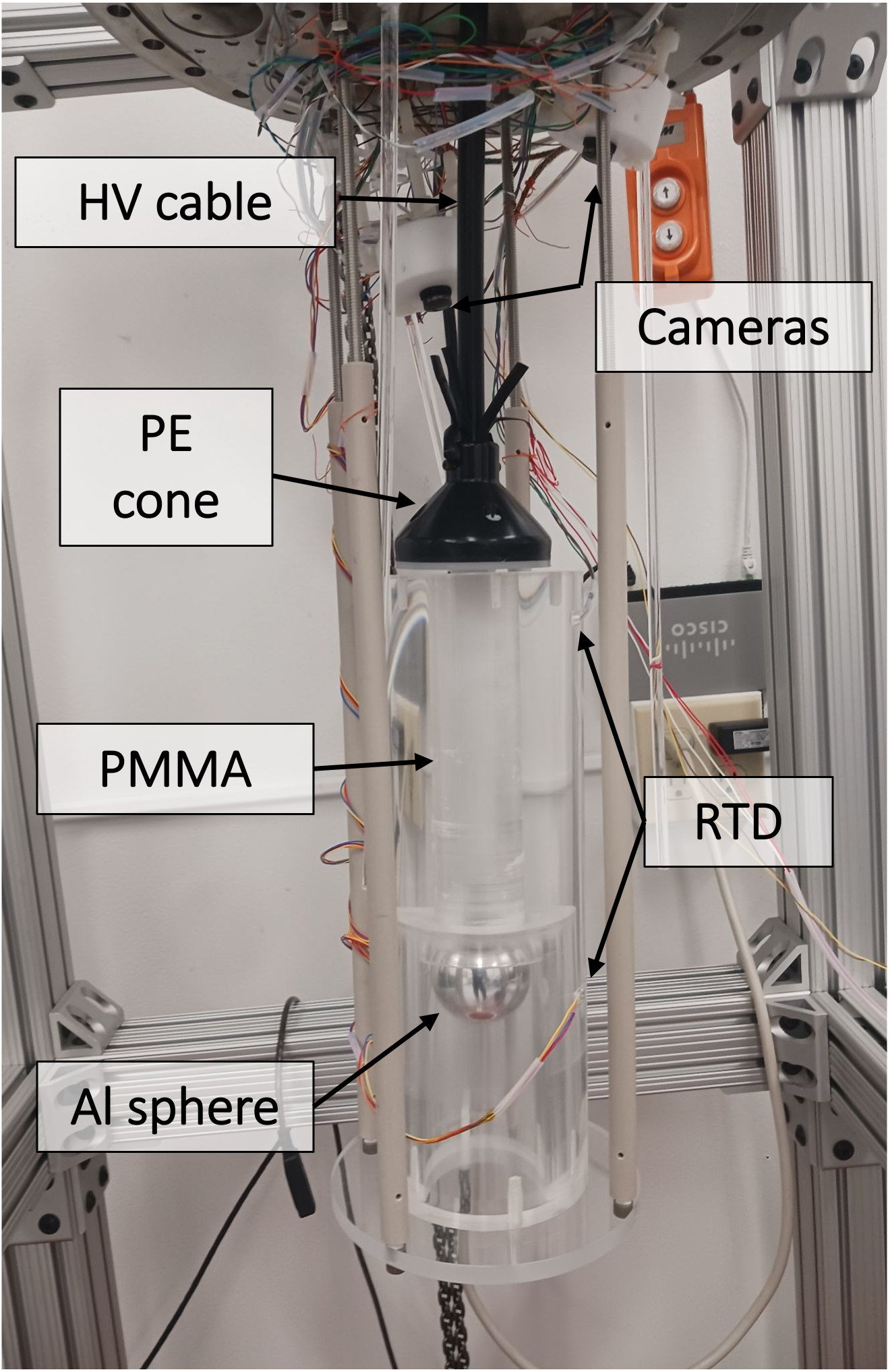}
        \caption{}
        \label{fig:hardware}
    \end{subfigure}
    \hspace{0.2\textwidth}
    \begin{subfigure}[b]{0.27\textwidth}
        \includegraphics[width=1.11\textwidth]{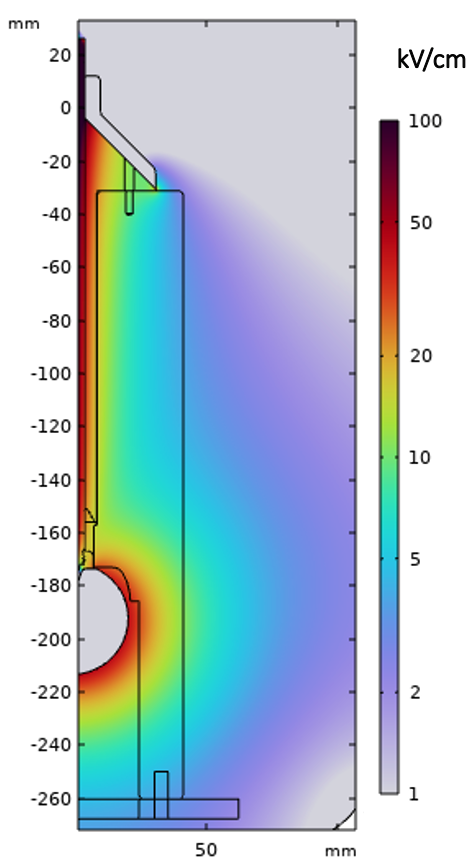}
        \caption{}
        \label{fig:comsol_1}
    \end{subfigure}
    \caption{Experimental test setup at UC Davis. Left: photograph of the hardware configuration. Right: 2D \texttt{COMSOL} simulation with the aluminum sphere biased at $-75$~kV.}
    \label{fig:setup}
\end{figure}

To investigate the operational margins of the DS--20k cathode HV system, a dedicated test campaign was performed at UC~Davis, where the DS--20k HV cable terminated with the PE--cone was operated inside a closed LAr system. The setup was designed to reproduce the DS--20k HV cathode plug in terms of local electric--field conditions, by inserting the PE--cone into a PMMA cylinder and connecting it to a 5 cm diameter aluminum sphere (Fig.~\ref{fig:setup}).

\noindent The entire setup was housed inside a SS dewar and instrumented to allow thermal and visual monitoring throughout cool--down and filling. The temperature was measured using two PT100 resistance temperature detectors (RTDs) mounted on the PMMA cylinder at two heights separated by $19.5$~cm. Electrical discharges and bubble formation were monitored using two Jinjiean B19 cryogenic cameras, with LED illumination employed only to enhance bubble visibility. Controlled and uniform cool--down was achieved using two heaters thermally coupled to a copper plate at the bottom of the dewar, whose temperature was monitored by an additional PT100 RTD.

A \texttt{COMSOL} simulation was used to validate the consistency of the test setup with the DS--20k cathode plug in terms of local electric--field conditions (Fig.~\ref{fig:comsol_1}). With the aluminum sphere biased at $-75$~kV, the electric fields in the critical regions were found to be comparable to those expected in DS--20k. Surface charge accumulation effects are not included in the simulation.

The test campaign aimed at reaching the nominal DS--20k operating voltage and up to $33\%$ above it, assessing the long--term stability of the system, validating HV ramp--up and ramp--down procedures, implementing voltage and current control and readout, monitoring the system optically, and establishing a reliable cool--down procedure.

%% file: TexFiles/Cryogenics.tex
\section{Cool--down and LAr Filling}
\label{sec:cryo}

Argon condensation was achieved using a Cryomech cryogenic refrigeration system. A key requirement of the cool--down procedure was to reduce the system temperature slowly and uniformly prior to argon condensation, in order to minimize thermal stress on the PMMA and PE components.

The SS vessel is narrow and double--insulated along its body, a configuration that favors temperature gradients and thermal stratification during cool--down. 
To mitigate these effects, bottom heaters thermally coupled to a copper plate were used to provide a thermal buffer against the cooling power, allowing controlled cooling in the gaseous phase and preventing abrupt temperature drops. The heated copper plate also enhanced convection in the lower region of the dewar, promoting a more uniform temperature distribution.
Temperature uniformity in the upper regions was supported by a Lihan CP5530 recirculation pump. 

Prior to filling, the entire system was evacuated to a base pressure of $7.6 \times 10^{-3}$~mbar. The dewar was then filled with industrial--grade argon (nominal purity $99.997\%$), for which the effective purity during the test was not independently measured. During the filling procedure, the system pressure was maintained close to atmospheric pressure ($\sim 1$~bar). Temperature, pressure, and argon flow rate were continuously monitored throughout the process.

The combined cool--down and LAr filling process required approximately $11$~days to reach a LAr level just above the stress--cone region, corresponding to a total LAr volume of about $20$~L. During this phase, the average cool--down rate was $\mathrm{d}T/\mathrm{d}t \simeq -0.5$~K/h, well below the conservative limit of $-4$~K/h adopted to prevent mechanical stress in the PMMA and PE components. This limit was defined based on experimental observations and operational experience with polymer materials under cryogenic conditions.
Throughout the cool--down, the temperature difference between the top and bottom PMMA RTDs was maintained $\Delta T_{\mathrm{top-bottom}} < 35$~K, corresponding to a vertical temperature gradient of less than $1.8$~K/cm between the two sensors.

%% file: TexFiles/High_Voltage.tex
\section{High--Voltage Operation and Results}
\label{sec:HV}

High voltage was supplied by a Heinzinger PNChp~100000--neg power supply ($V_{\max}=-100$~kV, $I_{\max}=1$~mA), identical to the unit foreseen for DS--20k.
The analog output of the power supply was interfaced with a National Instruments USB data acquisition device ($16$--bit resolution, $400$~kS/s), providing both control of the output voltage and continuous readout of both voltage and current. 

\begin{figure} [h]
    \centering
    \includegraphics[width=0.9\textwidth]{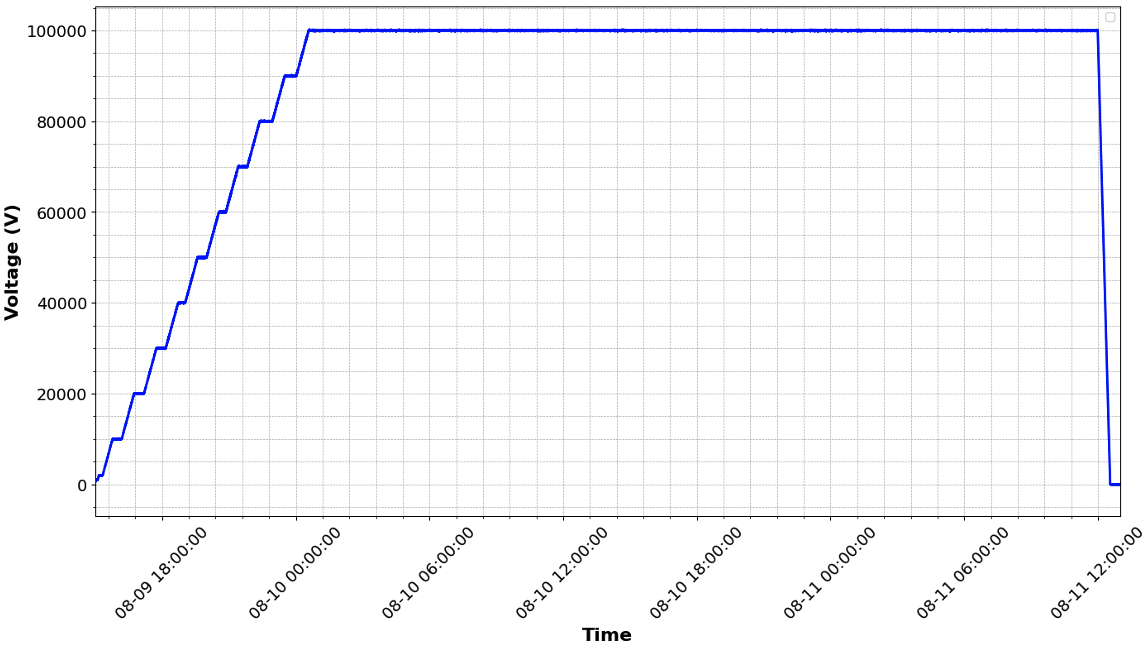}
    \caption{Measured output voltage during HV ramp--up to $-100$~kV at a rate of $5$~V/s and ramp--down to ground at $50$~V/s, as read out by the DAQ system over a $24$--h period.}
    \label{fig:voltage}
\end{figure}

The HV was ramped up to $-100$~kV at a rate of $5$~V/s in 10 kV steps, reaching the target voltage in approximately $10$~h. Ramp--down to ground was performed at $50$~V/s. The complete ramp--up and ramp--down sequence over a $24$--h period, as measured from the power--supply analog output by the DAQ system, is shown in Fig.~\ref{fig:voltage}. A long--term stability test was performed by operating the system at $-100$~kV for approximately $14$~days.

During HV operation, no voltage oscillations were observed within the $\sim1.5$~V readout resolution.
The measured current remained of order $\mathcal{O}(100~\mathrm{nA})$ during both ramp--up and ramp--down, with no current spikes indicative of electrical discharges. These observations show that the HV cable and PE--cone assembly can be operated up to $-100$~kV under the tested conditions.

From the cryogenic camera observations, neither light emission attributable to electrical discharges nor unambiguous bubble formation localized around the aluminum sphere was detected, the latter being hindered by the presence of multiple bubble sources within the dewar.

Post--test measurements of the HV cable showed capacitance and resistance values of $C = (376 \pm 4)$~pF and $R = (185 \pm 1)$~k$\Omega$, consistent within uncertainties with the pre--test measurements of $(380 \pm 4)$~pF and $(184 \pm 1)$~k$\Omega$, respectively. 
In addition, no mechanical damage attributable to thermal stress was observed on the PMMA or PE components of the HV cable and stress--cone assembly.

%% file: TexFiles/Conclusions.tex
\section{Conclusions}
\label{sec:conslusion}

In DS--20k, a nominal drift field of 200~V/cm requires a cathode bias of approximately $-75$~kV. To asses the operational margins of the cathode HV system, the DS--20k HV cable and stress--cone assembly were tested in a closed LAr system at UC~Davis under the same local electric--field conditions.
Following a controlled cool--down and filling with LAr, the system operated up to $-100$~kV for approximately $14$~days. No voltage oscillations were observed, the measured current remained of order $\mathcal{O}(100~\mathrm{nA})$. Post--test measurements showed capacitance and resistance of the HV cable consistent with pre--test values.
These results demonstrate that the HV cable and stress--cone assembly can be operated in LAr at voltages exceeding the nominal DS--20k cathode voltage, under controlled thermal and electrical conditions. This test provides technical and operational input for the design and operation of the DS--20k cathode HV supply and plug system. A next step foreseen for future tests is the integration of an argon purity monitor.